\documentclass[amsmath,amssymb]{revtex4}
\usepackage{graphicx}
\usepackage{epsfig}
\begin{document}
\title{Laboratory tests on dark energy}

\author{Christian Beck}

\address{School of Mathematical Sciences, Queen Mary, University of London,
Mile End Road, London E1 4NS, UK}

\email{c.beck@qmul.ac.uk}

\begin{abstract}
The physical nature of the currently observed dark energy in the
universe is completely unclear, and many different theoretical
models co-exist. Nevertheless, if dark energy is produced by
vacuum fluctuations then there is a chance to probe some of its
properties by simple laboratory tests based on Josephson
junctions. These electronic devices can be used to perform
`vacuum fluctuation spectroscopy', by directly measuring a noise
spectrum induced by vacuum fluctuations. One would expect to see
a cutoff near 1.7 THz in the measured power spectrum, provided
the new physics underlying dark energy couples to electric
charge. The effect exploited by the Josephson junction is a
subtile nonlinear mixing effect and has nothing to do with the
Casimir effect or other effects based on van der Waals forces. A
Josephson experiment of the suggested type will now be built, and
we should know the result within the next 3 years.
\end{abstract}

\maketitle

\section{Introduction---what is dark energy?}

It would be nice to start this paper with a clear definition of
what dark energy is and what it is good for. Unfortunately, the
answer to this question is completely unclear at the moment. What
is clear is that various astronomical observations
\cite{bennett03, spergel03} (supernovae,
CMB fluctuations, large-scale structure) provide rather
convincing evidence that around 73\% of the energy contents of
the universe is a rather homogeneous form of energy, so-called
`dark energy'. It behaves similar to a cosmological constant and
currently causes the universe to accelerate its expansion. Dark
energy may just be vacuum energy (with an equation of state
$w=p/\rho=-1$, where $p$ denotes the pressure and $\rho$ the
energy density). In that case its energy density $\rho$ is
constant and does not change with the expansion of the universe.
Or, $w$ may be just close to -1, in which case the dark energy
density evolves dynamically and changes with the expansion of the
universe. The remaining contents of the current universe is about
23\% dark matter and 4\% ordinary matter. With 96\% of the
universe being unknown stuff, there is enough room (and, indeed,
the need) for new theories. It seems that in order to construct a
convincing theory of dark energy that explains why it is there
and what role it plays in the universe one has to be open-minded
to new physics.

A large number of different theoretical models exist for dark
energy, but an entirely convincing theoretical breakthrough has
not yet been achieved. Popular models are based on quintessence
fields, phantom fields, quintom fields,
Born-Infeld quantum condensates, the
Chaplygin gas, fields with nonstandard kinetic terms,
to name just a few (see e.g.\
\cite{review, weinberg, caroll, pad} for reviews).
All of these approaches contain
`new physics' in one way or another, though at different levels.
However, it is clear that the number of possible dark energy
models that are based on new physics is infinite, and in that
sense many other models can be considered as well. Only
experiments will ultimately be able to confirm or refute the
various theoretical constructs.

\section{Dark energy from vacuum fluctuations}

A priori the simplest explanation for dark energy would be to
associate it with vacuum fluctuations that are allowed due to the
uncertainty relation. From quantum field theory it is well known
that virtual momentum fluctuations of a particle of mass $m$ and
spin $j$ formally produce vacuum energy given by
\begin{equation}
\rho_{vac}=\frac{1}{2}(-1)^{2j}(2j+1)\int_{-\infty}^{+\infty}
\frac{d^3k}{(2\pi)^3} \sqrt{{\bf k}^2+m^2} \label{vacontri}
\end{equation}
in units where $\hbar =c=1$. Here ${\bf k}$ represents the
momentum and the energy is given by $ E=\sqrt{{\bf k}^2+m^2}$.
Unfortunately, the integral is divergent: It formally yields
infinite vacuum energy density. Hence one has to find excuses why
this type of vacuum energy is not observable, or why only a very
small fraction of it survives.

A typical type of argument is that vacuum energy as given by
eq.~(\ref{vacontri}) is not gravitationally active, as long as
the theory is not coupled to gravity. This is more a belief
rather than a proved statement, and the statement looks a bit
fragile: Most particles in the standard model
of electroweak and strong interactions do have mass, so
they know what gravity is and hence it is not clear why their
vacuum energy should not be creating a gravitational effect if
their mass does. Another argument is that the vacuum energy might
be cancelled by some unknown symmetry, for example some type of
supersymmetry. For supersymmetric models, fermions with the same
mass as bosons generate vacuum energies of opposite sign, so all
vacuum energies add up to zero. Unfortunately, we know that we
currently live in a universe where supersymmetry is broken:
Nobody has ever observed bosonic electrons, having the same mass
as ordinary electrons. So this idea does not work either. We see
that in both cases the problem why we can't honestly get rid of
the vacuum energy comes from the masses of the particles, and
there is no theory of masses so far. In fact the Higgs mechanism,
which is supposed to create the particle masses in the standard
model of electroweak and strong interactions, creates an
additional (unacceptable) amount of vacuum energy, due to
symmetry breaking. So the only thing that we can say for sure is
that the problem of masses and vacuum energy in the current universe is
not yet fully understood.

The next, more pragmatic, step is then to introduce an upper
cutoff for the momentum $|{\bf k}|$ in the integral
(\ref{vacontri}). This leads to finite vacuum energy. But
typically, one would expect this cutoff to be given by the Planck
mass $m_{Pl}$, since we expect quantum field theory to be
replaced by a more sophisticated theory at this scale.
Unfortunately, this gives a vacuum energy density of the order
$m_{Pl}^4$---too large by a factor $10^{120}$ as compared to
current measurements of dark energy density. This is the famous
cosmological constant problem. The presently observed dark energy
density is more something of the order $m_\nu^4$, where $m_\nu$ is a
typical neutrino mass scale.

To explain the currently observed small dark energy density in
the universe, it seems one has to be open-minded to new physics.
This new physics can enter at various points. If vacuum
fluctuations (of whatever type) create dark energy, we basically
have two possibilities: Either, the dark energy is created by
vacuum fluctuations of ordinary particles of the standard model
(e.g. photons). The corresponding vacuum energy is given by the
integral (\ref{vacontri}), and the new physics should then
explain why there is a cancellation of the vacuum energy if the
momentum $|{\bf k}|$ exceeds something of the order $m_\nu$. Or,
there may be new types of vacuum fluctuations created by a new
type of dynamics underlying the cosmological constant which
intrinsically has a cutoff at around $m_\nu$. A priori, there is
no reason why this new dynamics of vacuum fluctuations should not
couple to electric charge. In fact, ultimately there is the need
to unify the quantum field theory underlying the standard model
with gravity, so the missing piece in this jigsaw should couple
to both. If the new dynamics of vacuum fluctuations (or the new
cancellation process of vacuum energy) couples to electric
charge, then there is a chance to see effects of this in
laboratory experiments on the earth \cite{mackey}, as we shall
work out in the following.

\section{Vacuum fluctuation spectroscopy with Josephson junctions}

It is well known that vacuum fluctuations produce measurable noise
in dissipative systems (see, e.g., \cite{gardiner} and references
therein). This noise has been experimentally verified in many
experiments. What is the deeper reason for the occurence of
quantum noise, e.g., in resistors? The basic principle is quite
easy to understand. Consider a dissipative system and two
canonically conjugated variables, say $x$ (position) and $p$
(momentum). If there are no external forces, then in the
long-term run the system will classically reach the stable fixed
point $p=0$, since all kinetic energy is dissipated. But quantum
mechanically, a state with $p=0$ would contradict the uncertainty
principle $\Delta x \Delta p =O(\hbar)$. Hence there must be
noise in the resistor that keeps the momentum going. This noise
acts as a fluctuating driving force and makes sure that the
momentum does not take on a fixed value, so that the uncertainty
principle is satisfied at all times.

Quantum noise is a dominant effect if the temperature is small.
It can be directly measured in experiments. Maybe one of the most
impressive experiments in this direction is the one done by Koch
et al. in the early eighties, based on resitively shunted
Josephson junctions \cite{koch80, koch82}. A Josephson junction
consists of two superconductors with an insulator sandwiched
inbetween. The behaviour of such a device can be modeled by the
following stochastic differential equation
\begin{equation}
 \frac{\hbar
C}{2e}\ddot{\delta}+\frac{\hbar}{2eR}\dot{\delta} +I_0\sin \delta
=I +I_N. \label{sto}
\end{equation}
Here $\delta$ is the phase difference across the junction, $R$ is
the shunt resistor, $C$ the capacitance of the junction, $I$ is
the mean current, $I_0$ the noise-free critical current, and
$I_N$ is the noise current. One can think of eq.~(\ref{sto}) as
formally describing a particle that moves in a tilted periodic
potential. The noise current $I_N$ produces some perturbations to
the angle of the tilt. There are two main sources of noise:
Thermal fluctuations and quantum fluctuations.

Koch et al. \cite{koch82} measured the noise spectrum at low
temperatures using four different
Josephson devices. They experimentally verified the following form of
the spectrum up to frequencies of $6\cdot 10^{11} Hz$:
\begin{eqnarray}
 S(\nu)&=&\frac{2h\nu}{R}\coth
\left( \frac{h\nu}{2kT} \right) \nonumber \\ &=&\frac{4h\nu}{R}
\left( \frac{1}{2}+\frac{1}{\exp (h\nu/kT)-1} \right)  .
\label{power}
\end{eqnarray}
The first term in  Equation  (\ref{power}) grows linear with the
frequency, as expected for the ground state of a quantum
mechanical oscillator. This term represents quantum noise and is
induced by zero-point fluctuations. The second term is ordinary
Bose-Einstein statistics and corresponds to thermal noise.

What is remarkable with the experiment of Koch et al. is the fact
that the quantum fluctuations produce a directly measurable
spectrum---many theoretical physicists thought (and
some still think)
that this is not possible. The experimentally
measured data of Koch et al. are shown in Fig.~1. In fact, noise
induced by quantum fluctuations plays an important role in any
resistor if the temperature is small enough. The Josephson
junction serves as a useful technical tool in this
context: Due
to a nonlinear mixing effect in the junction, one has the
experimental possibility to obtain measurements of very high
frequency noise. For recent theoretical work on the quantum
noise theory of Josephson junctions, see \cite{gardiner,levinson}.

The linear term in the spectrum is induced by zero-point
fluctuations and thus a consequence of the uncertainty relation.
Contrary to this, Jetzer and Straumann \cite{jetzer} have recently
expressed the view that the linear growth in the measured
spectrum is produced by van der Waals forces and that it has
nothing to do with zero-point fluctuations. But their view seems
to be at variance with the standard view of almost all experts in
the field of quantum noise theory (see \cite{gardiner} and
references therein). The standard view is that the linear term in
the measured spectrum is indeed a fingerprint of zero-point
fluctuations of the harmonic oscillators which model the
microscopic structure of the resistive element. Van der Waals
forces, as well as the Casimir effect, are different effects that
are relevant for other types of experimental situations (see,
e.g. \cite{casimir}), not for the noise in Josephson junctions.


\begin{figure}
\epsfig{file=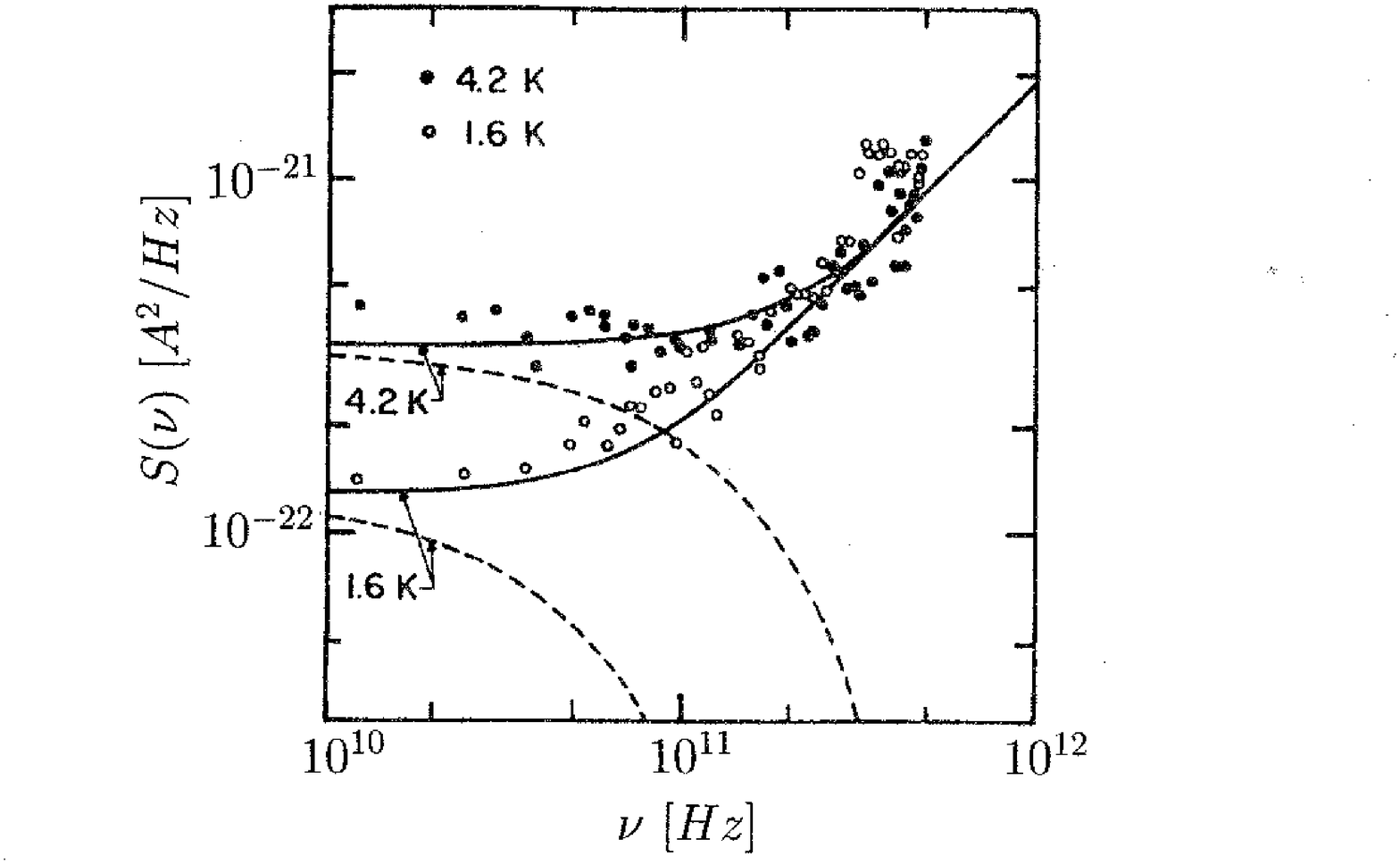, width=9cm, height=6.5cm}
\caption{Spectral density of current noise as measured in Koch et
al.'s experiment \cite{koch82} for two different temperatures. The
solid line takes into account the zero-point term and is given by
eq.~(\ref{power}) , whereas the dashed line is given by a purely
thermal noise spectrum, $(4h\nu/R)(exp(h\nu/kT)-1)^{-1}$.}

\end{figure}

Since zero-point fluctuations produce experimentally measurable
effects in Josephson junctions, it is natural to conjecture that
the energy density associated with the underlying primary fluctuations has
physical meaning as well: It is a prime candidate for dark
energy, being isotropically distributed and temperature
independent. There is vacuum energy associated with the measured
data in Fig.~1, and it cannot be easily discussed away.

Of course, one has to be careful in the interpretation of the
experimental data in Fig.~1. What is really measured with the
Josephson junction are real currents produced by real electrons.
These currents behave very much in a classical way, however, they
are {\em induced} by zero-point fluctuations. What type of
zero-point fluctuations induce the measured currents in the first
place is not clear from a theoretical point of view. It could be
just ordinary vacuum fluctuations (of QED type), but there is
also the possibility of other, new types of vacuum fluctuations,
which could potentially underly dark energy and produce a
measurable effect in the Josephson junction. What is clear is
that if there were no zero-point fluctuations in the first place,
then there would also be no linear term in the measured spectrum.
Moreover, if the vacuum fluctuations inducing the noise in the
junction had a cutoff at some frequency $\nu_{max}$, and would
fall quiet above that, the same effect would be observed for the
induced physical noise spectrum. In any case, if such a cutoff
exists then it would be the {\em new physics} underlying the
cutoff mechanism that makes the system couple to gravity and thus
make the corrseponding vacuum energy physically relevant. This is
the basic idea underlying the recent suggestion of laboratory
tests on dark energy
---checking whether there is a cutoff or not in an extended
new version of the Koch experiment
\cite{mackey, nature}.

\section{Estimating a cosmological cutoff frequency}

%

Already Planck \cite{planck14} and Nernst \cite{nernst16}
formally included zero-point terms in their work. Their formulas yield
for the energy density of a collection of oscillators of frequency $\nu$
\begin{eqnarray}
    \rho(\nu,T) &=& \frac{8 \pi \nu^2}{c^3}
    \left [ \frac{1}{2}h \nu  + \frac {h \nu }{\exp(h \nu /kT)-1} \right ] \nonumber \\
    &=& \frac{4   \pi  h \nu^3}{c^3}
    \coth \left ( \frac{h\nu}{2kT} \right ). \label{four}
     \label{nu-spectrum}
\end{eqnarray}
This just corresponds to the measured spectrum in the Josephson
junction experiment, up to a prefactor depending on $R$. We may
split eq.~(\ref{four}) as
\begin{equation}
\rho(\nu,T)=\rho_{vac}(\nu) +\rho_{rad}(\nu,T), \label{4}
\end{equation}
where
\begin{equation}
\rho_{vac}(\nu)=\frac{4\pi h\nu^3}{c^3} \label{vac}
\end{equation}
is induced by zero-point fluctuations, and
\begin{equation}
\rho_{rad}(\nu , T)=\frac{8\pi h \nu^3}{c^3}
\frac{1}{\exp{(h\nu/kT)}-1}. \label{rad}
\end{equation}
corresponds to the radiation energy density generated by photons
of energy $h\nu$. Integration of eq.~(\ref{vac}) up to some cutoff
$\nu_c$ yields
\begin{equation}
     \int_0^{\nu_c}  \rho_{vac}(\nu) d\nu =
    \frac {4   \pi  h}{c^3} \int_0^{\nu_c} \nu^3
    d\nu = \frac {   \pi  h}{c^3}  \nu_c^4.
    \label{total-2}
\end{equation}
This is equivalent to eq.~(\ref{vacontri}) with $m=0$.
Integration of
eq.~(\ref{rad}) over all frequencies yields the well-known
Stefan-Boltzmann law
\begin{equation}
\int_0^\infty \rho_{rad}(\nu,T)d\nu =
\frac{\pi^2k^4}{15\hbar^3c^3} T^4.
\end{equation}
Assuming that vacuum fluctuations (of whatever type) are
responsible for dark energy, the necessary cutoff frequency
follows from the current astronomical estimates of dark energy
density \cite{bennett03,spergel03}:
\begin{equation}
    \rho_{dark} = 0.73 \rho_c  = (3.9 \pm 0.4) \quad \mbox{GeV/m}^3
\end{equation}
Here $\rho_c$ denotes the critical density of a flat universe.
From
\begin{equation}
    \frac{   \pi  h}{c^3}  \nu_c^4 \simeq \rho_{dark}
\end{equation}
we obtain
\begin{equation}
    \nu_c \simeq (1.69 \pm 0.05) \times 10^{12} \quad \mbox{Hz}.
\label{cutoff}
\end{equation}

So if vacuum fluctuations underly dark energy, and if these
vacuum fluctuations drive the corresponding quantum oscillators
of frequency $\nu$ in the resistive element, one would expect to
see a cutoff near $\nu_c$ in the measured spectrum of the
Josephson junction experiment. Because otherwise the corresponding
vacuum energy density would exceed the currently measured dark
energy density. The frequency $\nu_c$ is about 3 times higher than
the largest frequency reached in Koch et al.'s experiment of 1982.
Future experiments, based on new Josephson junction technology,
will be able to reach this higher frequency (see last section).

Suppose a cutoff near $\nu_c$ is observed in a future
experiment. Then this would represent new physics and
the consequences would be far reaching. Frampton \cite{frampton} has
shown that the observation of such a cutoff could even shake some
of the assumptions underlying string theory, leading to the
possible demise of the so-called string landscape. His
argument is based on the fact that if the cutoff is seen in the
Josephson experiment then this implies that the dark energy field
interacts with the electromagnetic field---which leads to a potential
problem for string theory if the physical vacuum,
as it is usually assumed, decays by a 1st
order phase transition:
A small cosmological constant
as generated in this context would have decayed to zero by now,
contradicting the
fact that we do see dark energy right now. In any case, it is
interesting that for the first time there seems to be an
experiment that can
check some of the assumptions underlying the string landscape.

\section{New types of vacuum fluctuations?}

As said before: If we allow for the possibility of new physics
then there are many possible models of dark energy. Let us here
consider the possibility that dark energy is produced by {\em
new} types of vacuum fluctuations with a suitable cutoff. The
model was introduced in detail in \cite{prd}, here we just sketch
the main idea.

We start from a homogeneous self-interacting scalar field
$\varphi$ with potential $V(\varphi)$.
Most dark energy models are just formulated in a
classical setting, but for our approach in terms of vacuum
fluctuations we need of course to proceed to a second-quantized
theory. We second-quantize our scalar field using the Parisi-Wu
approach of stochastic quantization \cite{parisi}, which is a
convenient and useful method for our approach. The
second-quantized field $\varphi$ then obeys a stochastic differential equation
of the form
\begin{equation}
\frac{\partial}{\partial s}\varphi =\ddot{\varphi}
+3H\dot{\varphi} +V'(\varphi) +L(s,t), \label{sto2}
\end{equation}
where $H$ is the Hubble parameter, $t$ is physical time, $s$ is
fictitious time (just a formal coordinate to do quantization) and
$L(s,t)$ is Gaussian white noise, $\delta$-correlated both in $s$
and $t$. The fictitious time $s$ is introduced as a formal
tool for 2nd quantization, it has dimensions $GeV^{-2}$. Quantum
mechanical expectations can be calculated as expectations of the
above stochastic process for $s \to \infty$. The advantage of the
Parisi-Wu approach is that it is very easy and natural to
introduce cutoffs in this formulation --- by far easier than in
the canonical field quantization approach. The simplest way to
introduce a cutoff is by making $t$ and $s$ discrete (as in any
numerical simulation of a stochastic process). Hence we write
\begin{eqnarray}
s &=& n\tau \\ t &=& i \delta ,
\end{eqnarray}
where $n$ and $i$ are integers and $\tau$ is a fictitious time
lattice constant, $\delta$ is a physical time lattice constant.
Note that the uncertainty relation $\Delta E \Delta t =O(\hbar)$
always implies an effective lattice constant $\Delta t$ for a
given finite energy $\Delta E$.
We also introduce a dimensionless field variable $\Phi_n^i$ by
writing $\varphi_n^i=\Phi_n^i p_{max}$, where $p_{max}$ is some
(so far) arbitrary energy scale. The above scalar field dynamics
is equivalent to a discrete dynamical system of the form
\begin{equation}
\Phi_{n+1}^i=(1-\alpha)T(\Phi_n^i)+\frac{3}{2}H\delta \alpha
(\Phi_n^i-\Phi_n^{i-1})+\frac{\alpha}{2}(\Phi_n^{i+1}+\Phi_n^{i-1})
+ \tau\cdot noise, \label{dyn}
\end{equation}
where the local map $T$ is given by
\begin{equation}
T(\Phi )=\Phi
+\frac{\tau}{p_{max}(1-\alpha)}V'(p_{max}\Phi)\label{map}
\end{equation}
and $\alpha$ is defined by
\begin{equation}
\alpha:=\frac{2\tau}{\delta^2}.
\end{equation}
For old universes, one can neglect the term proportional to $H$,
obtaining
\begin{equation}
\Phi_{n+1}^i=(1-\alpha)T(\Phi_n^i)+\frac{\alpha}{2}(\Phi_n^{i+1}+\Phi_n^{i-1})
+\tau \cdot noise \label{sym}
\end{equation}
We now want to construct a field that basically manifests itself as
noise:
Rather than evolving smoothly it should exhibit strongly
fluctuating behavior, so that we may be able to interpret its
rapidly fluctuating behaviour in terms of
vacuum fluctuations, and possibly in terms of measurable noise
in the Josephson junction. As
a distinguished example of a $\varphi^4$-theory generating such
behaviour, let us consider the map
\begin{equation}
\Phi_{n+1}=T_{-3}(\Phi_n)=-4\Phi_n^3+3\Phi_n
\end{equation}
on the interval $\Phi\in [-1,1]$. $T_{-3}$ is the negative
third-order Tchebyscheff map, a standard example of a map
exhibiting strongly chaotic behaviour. It is conjugated to a
Bernoulli shift, thus generating the strongest possible chaotic
behaviour possible for a smooth low-dimensional deterministic
dynamical system \cite{spatio}. The corresponding potential is given
by
\begin{equation}
V_{-3}(\varphi)=\frac{1-\alpha}{\tau}\left\{
\varphi^2-\frac{1}{p_{max}^2} \varphi^4\right\}+const, \label{16}
\end{equation}
or, in terms of the dimensionless field $\Phi$,
\begin{equation}
V_{-3}(\varphi)=\frac{1-\alpha}{\tau} p_{max}^2 ( \Phi^2 -\Phi^4)
+ const. \label{17}
\end{equation}
The important point is that starting from this potential we obtain
by second quantization a field $\varphi$ that rapidly fluctuates
on some finite interval, choosing initially
$\varphi_0\in [-p_{max},p_{max}]$. Since these chaotic
fluctuations are bounded, there is a natural cutoff.

The idea is now that the expectation of the potential of this
chaotic field (plus possibly kinetic terms) underly the measured
dark energy density in the universe. Expectations $\langle \cdots
\rangle$ can be easily numerically determined by iterating the
dynamics (\ref{sym}) for random initial conditions. One has
\begin{equation}
\langle V_{-3}(\varphi)\rangle =\frac{1-\alpha}{\tau} p_{max}^2 (
\langle \Phi^2\rangle  -\langle \Phi^4\rangle ) + const,
\end{equation}
which for $\alpha=0$ can be analytically evaluated \cite{spatio} to
give
\begin{equation}
\langle V_{-3} (\varphi)
\rangle=\frac{1}{8}\frac{p_{max}^2}{\tau} +const.
\end{equation}

To reproduce the currently measured dark energy, we only need to
fix the ratio of the parameters $\tau$ and $p_{max}$ as
\begin{equation}
\frac{p_{max}^2}{\tau} \sim \rho_\Lambda \sim m_\nu^4
\end{equation}
This is the simplest model of noise-like vacuum fluctuations with
a suitable finite cutoff one can think of, a chaotic scalar field
theory underlying the cosmological constant. It is easy to show
\cite{prd} that for $\alpha =0$ the equation of state of this
field is $w=-1$. For small $\alpha$, it is close to $w=-1$. In
this model, there is no reason why the deterministic chaotic noise
represented by this rapidly fluctuating field should not be able
to influence electric charges. Hence these chaotic fluctuations
may well induce a measurable noise spectrum in Josephson
junctions.


\section{A new quantum noise experiment}

In \cite{mackey} we suggested to repeat the experiment of Koch et
al. with new types of Josephson junctions that are capable of
reaching higher frequencies. This new experiment will now be
built, the grant has just been allocated \cite{experiment}.
Warburton, Barber, and Blamire are planning two different
versions of the experiment. One is based on nitride junctions,
the other one on cuprate junctions. The maximum frequency that can
be reached is determined by the gap energy of the Josephson
junction under consideration, and the above materials provide the
possibility to reach the cosmologically interesting frequency of
$\nu_c \approx 1.7$ THz and even exceed it. The new technology
suggested by Warburton et al.\ is more sophisticated than that of
conventional Niobium-based Josephson junctions. With conventional
Niobium-based junctions one would probably be only able to reach
1.5 THz in the measurements since their gap energy is too small.
By performing experiments on both the nitrides and the cuprates
there will be two independent high frequency measurements of the
quantum noise spectrum in two very different material systems. So
in about 3 years time we should know whether there is any unusual
behaviour of zero-point fluctuations near $\nu_c$, which could
possibly be related to dark energy. Whatever the outcome of this
new experiment, the result will be interesting: If a cutoff is
observed it will revolutionize our understanding of dark energy.
If a cutoff is not observed, it will show that the vacuum
fluctuations measured with the Josephson junction are definitely
not gravitationally active.

\end{document}